\documentclass{emulateapj}
\usepackage{graphicx}
\usepackage{epstopdf}

\newcommand{\Mjup}{\mbox{$M_\mathrm{Jup}$}}
\newcommand{\Msun}{\mbox{$M_{\odot}$}}
\newcommand{\Mearth}{\mbox{$M_{\earth}$}}

\begin{document}
\title{An ALMA Constraint on the GSC~6214-210~B Circum-Substellar Accretion Disk Mass}
\author{Brendan P. Bowler,\altaffilmark{1,6} 
Sean M. Andrews,\altaffilmark{2} 
Adam L. Kraus,\altaffilmark{3} 
Michael J. Ireland,\altaffilmark{4}
Gregory Herczeg,\altaffilmark{5}
Luca Ricci,\altaffilmark{2}
John Carpenter,\altaffilmark{1}
Michael E. Brown\altaffilmark{1}
\\ }
\email{bpbowler@caltech.edu}

\altaffiltext{1}{California Institute of Technology, 1200 E. California Blvd., Pasadena, CA 91125, USA}
\altaffiltext{2}{Harvard-Smithsonian Center for Astrophysics, 60 Garden Street, Cambridge, MA 02138, USA}
\altaffiltext{3}{Department of Astronomy, The University of Texas at Austin, Austin, TX 78712, USA}
\altaffiltext{4}{Research School of Astronomy \& Astrophysics, Australian National University, Canberra ACT 2611, Australia}
\altaffiltext{5}{Kavli Institute for Astronomy and Astrophysics, Peking University, Yi He Yuan Lu 5, Haidian Qu, Beijing 100871, China}
\altaffiltext{6}{Caltech Joint Center for Planetary Astronomy Fellow.}

\begin{abstract}
We present Atacama Large Millimeter/submillimeter Array (ALMA) observations of GSC~6214-210~A and B, 
a solar-mass member of the 5--10~Myr Upper Scorpius association with a 15~$\pm$~2~\Mjup \ companion  orbiting 
at $\approx$330 AU (2$\farcs$2).
Previous photometry and spectroscopy spanning 0.3--5~$\mu$m revealed optical and thermal 
excess as well as strong H$\alpha$ and Pa~$\beta$ emission originating from a circum-substellar 
accretion disk around GSC~6214-210~B, making it the lowest mass companion with 
unambiguous evidence of a subdisk.
Despite ALMA's unprecedented sensitivity and angular resolution, 
neither component was detected in
our 880~$\mu$m (341~GHz) continuum observations down to a 3-$\sigma$ limit of 0.22~mJy/beam.
The corresponding constraints on the dust mass and total mass are $<$0.15~$\Mearth$ and 
$<$0.05~\Mjup, respectively, or $<$0.003\% and $<$0.3\% of the mass of GSC~6214-210~B itself
assuming a 100:1 gas-to-dust ratio and characteristic dust temperature of 10--20~K.   
If the host star possesses a putative circum-stellar disk then at most it is a meager 0.0015\% 
of the primary mass, implying that giant planet formation has certainly ceased in this system.
Considering these limits and its current accretion rate, GSC~6214-210~B appears to be at the end 
stages of assembly and is not expected to gain any appreciable mass over the next few Myr.
\end{abstract}

\section{Introduction}{\label{sec:intro}}

Giant planets are common products of protoplanetary disk evolution but many of the details of their formation 
remain obscure and untested. 
Core accretion is the dominant formation channel at small separations of $\lesssim$10~AU 
  (\citealt{Pollack:1996jp}; \citealt{Alibert:2005ee}), while disk instability (\citealt{Boss:1997di})
  and turbulent fragmentation (\citealt{Bate:2009br}) operate on wider orbits at tens to hundreds of AU.
In these outer regions ($\sim$10--300~AU), high-contrast direct imaging surveys are revealing that the tail end of the
substellar companion mass function is likely to be continuous (\citealt{Brandt:2014cw}), 
extends down to at least $\approx$5~\Mjup \ (\citealt{Rameau:2013ds}; \citealt{Kuzuhara:2013jz}), and
overlaps with the most massive planets like HR~8799~bcde (\citealt{Marois:2008ei}; \citealt{Marois:2010gpa})
and $\beta$~Pic~b (\citealt{Lagrange:2009hq}), which probably formed in massive protoplanetary disks.

In all formation scenarios, giant planets accrete most of their mass from circum-planetary disks, which naturally form around 
protoplanets in an analogous way to circumstellar disks around protostars (\citealt{Vorobyov:2010bs}; \citealt{Ward:2010fc}).  
Circum-planetary disks regulate angular momentum transport and accretion as protoplanets grow, ultimately determining 
the final masses of giant planets (\citealt{Quillen:1998dw}; \citealt{Lubow:1999dk}; \citealt{Canup:2002ii}; 
\citealt{Zhu:2015fr}).  
At the earliest stages of giant planet assembly ($\lesssim$3~Myr), 
accretion shocks from circum-planetary disks onto protoplanets help to define planets' initial entropy levels and which
evolutionary pathways (hot, warm, or cold start models) they will follow (\citealt{Marley:2007bf}).
Moreover, the formation, bulk composition, and orbits of moons are also governed by properties of circum-planetary disks (\citealt{Heller:2014fy}).
Physically, these disks are expected to be quite thick ($h$/$r$ $>$ 0.2), possess little flaring 
compared to their circum-stellar analogs, and be tidally truncated to $\sim$1/3 $R_\mathrm{Hill}$ from  gravitational influence 
of the host star (\citealt{Ayliffe:2009bx}; \citealt{Martin:2011hc}; \citealt{Shabram:2013dm}).

This epoch of planet growth is challenging to directly study because of the small angular 
resolution and high contrasts required.
The possible discovery of extended thermal excess emission from the candidate protoplanets 
LkCa~15~b (\citealt{Kraus:2012gk}) and HD~100546~b (\citealt{Quanz:2013ii}) may represent the first direct detections of 
circum-planetary disks, though the masses and immediate environments of these embedded companions are still 
poorly constrained (\citealt{Isella:2014fz}).
The remnants of circum-planetary disks are evident from the prograde, regular satellites orbiting gas giants in our Solar System
as well as the (possibly transient) ring 
systems surrounding Saturn, the substellar companion 1SWASP~J140747.93--394542.6 
(\citealt{Mamajek:2012dn}; \citealt{Kenworthy:2014ga}), and perhaps Fomalhaut~b (\citealt{Kalas:2008cs}).

\begin{figure*}
  \vskip -.3 in
  \hskip 0 in
  \resizebox{6.8in}{!}{\includegraphics{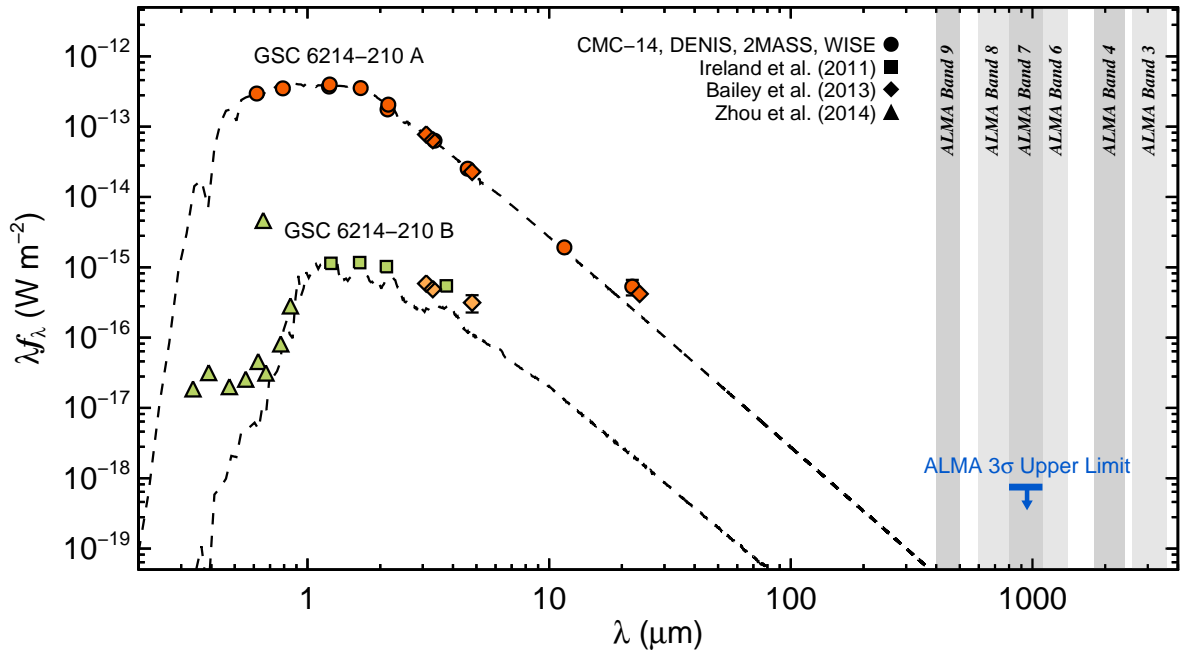}}
  \vskip -1.3 in
  \caption{Spectral energy distributions of GSC 6214-210 A (red) and B (green).  GSC 6214-210~B
  exhibits optical and thermal excess along with extraordinarily strong H$\alpha$ emission.  
  The unresolved pair shows slight excess beyond 12~$\mu$m that could originate
  from the circum-substellar disk around GSC 6214-210~B and/or a low mass disk around the primary.
  Our ALMA 880~$\mu$m 3-$\sigma$ upper limit (blue) rules out substantial disks from both
  components.    Photometry for
  the primary are from CMC-14 ($r'$), DENIS ($IJK$), 2MASS ($JHK$), $WISE$ 
  (3.4, 4.6, 12, and 22~$\mu$m), and $Spitzer$/MIPS (24~$\mu$m; \citealt{Bailey:2013gl}).
  Note that the 3.1~$\mu$m, 3.3~$\mu$m, and $M'$ photometry of GSC 6214-210 A from \citet{Bailey:2013gl}
  are inferred rather than independently measured.  Photometry for GSC 6214-210 B are from 
  \citet[$JHKL'$]{Ireland:2011id}, \citet[3.1~$\mu$m, 3.3~$\mu$m, $M'$]{Bailey:2013gl}, 
  and \citet[0.33--0.85~$\mu$m]{Zhou:2014ct}.  The 
  BT-Settl 4200~K/4.0~dex and 2400~K/4.5~dex models (dashed lines; \citealt{Allard:2010wp}) 
  scaled to the $J$-band photometry are shown for comparison.  \label{fig:sed} } 
\end{figure*}

Over the past decade a population of young ($\lesssim$10~Myr) planetary-mass companions on ultra-wide orbits ($>$100~AU) has been uncovered with direct
imaging (see Table~1 of \citealt{Bowler:2014dk}).  Many of these
possess their own accretion disks, offering unique opportunities to study the structure, diversity, and evolution 
of circum-planetary disks.
Recently \citet{Kraus:2015fx} presented 1.3~mm Atacama Large Millimeter/submillimeter Array (ALMA) 
continuum observations of FW~Tau~C, a young ($\approx$2~Myr), wide-separation (330~AU) companion 
with a possible mass as low as 6--14~\Mjup \ (\citealt{Kraus:2014tl}).
The clear detection at mm wavelengths implies a dust mass of 1--2~\Mearth, sufficient to form 
a system of rocky moons.  However, the high veiling and possible edge-on architecture of 
the this accretion disk makes the mass of FW~Tau~C highly uncertain and possibly well above the planetary
regime (\citealt{Bowler:2014dk}).

Here we present ALMA 880~$\mu$m continuum observations of 
GSC~6214-210~B, a $\approx$15~\Mjup \ companion orbiting the young (5--10~Myr) 
Sun-like star GSC~6214-210~A in Upper Scorpius (USco) at 330~AU.  With a mass at the deuterium-burning limit
and abundant evidence of ongoing accretion, GSC~6214-210~B is the lowest-mass 
companion with a circum-substellar disk currently known.
ALMA's unprecedented sensitivity and angular resolution at sub-mm and mm wavelengths
presents a unique opportunity to characterize a sub-disk at the boundary between the
brown dwarf and planetary mass regimes.  Additionally, ALMA observations provide
important contextual information about any circum-stellar disk surrounding 
GSC~6214-210~A, or lack thereof in case of a non-detection.

\section{The GSC 6214-210 AB System}{\label{sec:gsc}}

GSC 6214-210 is a well-established member of the USco star-forming region
(\citealt{Preibisch:1998ur}; \citealt{Ireland:2011id}; \citealt{Luhman:2012hj}).  Low-mass 
evolutionary models point to an age of $\approx$5~Myr for this complex, though more recent 
turn-off ages suggest it is somewhat older ($\approx$10~Myr; \citealt{Pecaut:2012gp}).  
Based on its spectral type (K5~$\pm$~1) and modest reddening ($A_V$$\sim$0.6~mag), 
evolutionary models imply a mass of 0.9~$\pm$~0.1~\Msun \ for the primary 
(\citealt{Baraffe:1998ux}; \citealt{Bowler:2011gw}).  GSC 6214-210 is an evolved Class~III object 
with no signs of a substantial disk out to 12~$\mu$m.   

The low-mass companion to GSC 6214-210 was first identified at 2$\farcs$2 (330~AU) 
from Keck/AO imaging by \citet{Kraus:2008bh} and confirmed to be comoving with the 
primary by \citet{Ireland:2011id}.  
Evolutionary models imply a mass of 15~$\pm$~2~\Mjup \  from its age (5--10~Myr) and luminosity 
(log~$L_\mathrm{bol}$/$L_{\odot}$ = --3.1~$\pm$~0.1~dex).
\citet{Bowler:2011gw, Bowler:2014dk} found a near-infrared spectral type of M9.5~$\pm$~1 
for GSC~6214-210~B as well as strong Pa$\beta$ emission at 1.282~$\mu$m ($EW$=--11.2~\AA), 
indicating vigorous accretion from a circum-substellar disk.  
Recently, 1.0--2.4~$\mu$m spectroscopy by \citet{Lachapelle:2015cx} also revealed 
Pa$\beta$ emission as well as weak Br~$\gamma$ emission at 2.166~$\mu$m.
Optical imaging with $HST$ by \citet{Zhou:2014ct} uncovered extraordinarily strong H$\alpha$ emission 
($EW$$\sim$--1600~\AA) and continuum excess shortward of $\sim$6000~\AA, corresponding 
to an accretion rate of $\dot{M}$$\sim$10$^{-10.8}$~\Msun \ yr$^{-1}$.
3--5~$\mu$m imaging by \citet{Ireland:2011id} and \citet{Bailey:2013gl} supports the presence of a 
warm disk from thermal excess emission above the photospheric level.
Unresolved 22~$\mu$m photometry of GSC~6214-210~AB from $WISE$ and 24~$\mu$m photometry from $Spitzer$
(\citealt{Bailey:2013gl}) reveals a slight (but significant) 
excess, which could be from an evolved disk around the primary and/or a massive disk 
surrounding the companion (Figure~\ref{fig:sed}).
Synthesized $W4$-band photometry of the 4200~K/4.0~dex and 2400~K/4.5~dex BT-Settl models in Figure~\ref{fig:sed}
implies a range of flux densities for the companion between 9.92$\times$10$^{-20}$ W~m$^{-2}$~$\mu$m$^{-1}$
(or 2.19$\times$10$^{-18}$ W~m$^{-2}$ at 22~$\mu$m, assuming no excess) to 
1.2~$\pm$~0.6$\times$10$^{-17}$ W~m$^{-2}$~$\mu$m$^{-1}$ 
(or 2.6~$\pm$~1.3$\times$10$^{-16}$ W~m$^{-2}$ at 22~$\mu$m, assuming the excess is entirely attributed to the companion).

\begin{figure}
  \vskip -.8 in
  \hskip -2 in
  \resizebox{6.1in}{!}{\includegraphics{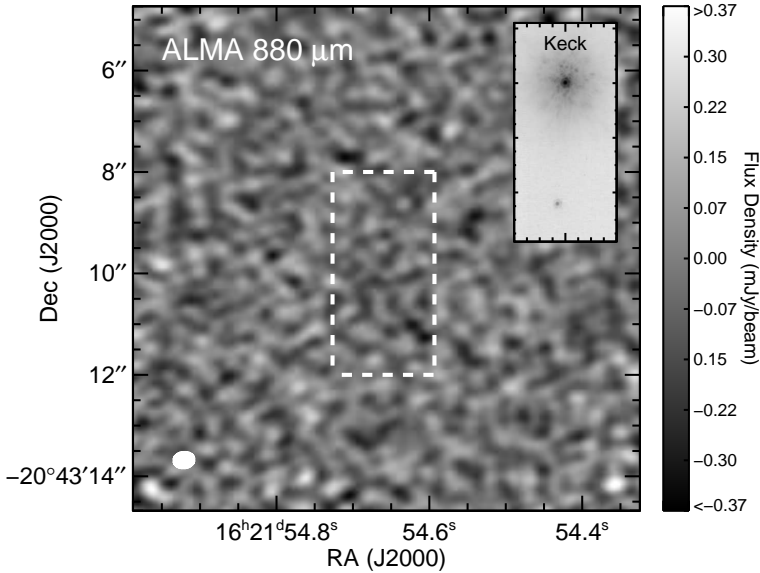}}
  \vskip -.5 in
  \caption{ALMA 880~$\mu$m continuum image centered on GSC~6214-210~AB.  The inset shows 
  a 2$''$$\times$4$''$ $K'$-band Keck/NIRC2 adaptive optics image of the pair from \citet{Ireland:2011id}.
  The same region is depicted as a dashed box in the ALMA image.  Neither component is detected with a 
  3-$\sigma$ upper limit of 0.22~mJy (assuming a point source).  
  The field of view is 10$''$$\times$10$''$ and the 0$\farcs$3$\times$0$\farcs$4 beam FWHM is 
  shown as a white ellipse in the bottom left corner.  The image has been corrected for the 
  nonuniform sensitivity of the primary beam across the field of view, and the color stretch spans
  $\pm$5$\sigma$ ($\pm$0.37~mJy/beam).
  \label{fig:rms} } 
\end{figure}

\section{ALMA Observations}

GSC~6214-210~AB ($\alpha$=16:21:54.67, $\delta$=--20:43:09.2) was observed with ALMA in Band 7 
at 880~$\mu$m (341 GHz) during Cycle 2 Early Science operations on UT 2014 June 30.  
31 12-meter antennas targeted this system with the dual-polarization setup in four spectral windows centered 
at 334.0, 336.0, 346.0, and 347.8~GHz.  The 346.0~GHz window encompases the CO ($J$=3-2) line
and was sampled in 1920 channels each with a width of 0.98~MHz (frequency division correlator mode), 
while the other three continuum windows comprised 128 channels at 15.6~MHz each (time division correlator mode).
The total on-source integration time was 11.6 minutes and the total effective bandwidth was 7.7~GHz.
The maximum array baseline reached 650~m, resulting in a  synthesized beam 
FWHM of 0$\farcs$32$\times$0$\farcs$41 at a position angle of 96.3$^{\circ}$ and a field of
view spanning 18$''$$\times$18$''$.

The visibility data reduction and image reconstruction were carried out with \texttt{CASA} version 4.2.2.
The primary calibrator was the nearby quasar QSO~J1625--2527, which was targeted intermittently during the  
science observations.
No line emission was detected in the 346.0~GHz data so this window was incorporated with the continuum regions
to create an averaged continuum image.  
No detections are present above the background rms level of
0.074~mJy/beam as measured in the central 3$\farcs$2$\times$3$\farcs$2 region. 
Finally, the continuum image was corrected for the non-uniform sensitivity
across the field of view by dividing by the primary beam response.  The
result is shown in Figure~\ref{fig:rms}.

\section{Results}

There is no significant emission above the background level (0.074~mJy/beam rms) at the location 
of GSC~6214-210~A or its companion, implying a 3-$\sigma$ upper limit of 0.22 mJy (assuming a point source).  
With a spatial resolution of $\approx$0$\farcs$4, both components would be easily separated 
in our observations.  Assuming the continuum emission is optically thin, an upper limit 
on the dust mass ($M_\mathrm{dust}$) can be derived 
from our ALMA flux density constraint, the distance to the system (145~pc), a characteristic 
isothermal disk temperature ($T_C$), and a dust opacity 
($\kappa_\nu$) following \citet{Hildebrand:1983tm}.  
For consistency with previous work we use the frequency-dependent dust opacity relationship 
$\kappa_\nu$ = 10($\nu$/10$^{12}$ Hz) cm$^2$ g$^{-1}$ from \citet{Beckwith:1990hj}.  
At 880~$\mu$m, the dust opacity $\kappa_\mathrm{880 \mu m}$ is 3.4 cm$^2$ g$^{-1}$.

Following \citet{Andrews:2013ku}, we adopt the dust temperature-stellar luminosity relationship
$T_C$ = 25 ($L$/$L_{\odot}$)$^{1/4}$ for the primary GSC~6214-210~A.
This yields $T_C$=20~K using the luminosity of log~$L$/$L_\mathrm{Bol}$ = --0.42~$\pm$~0.08~dex
from \citet{Bowler:2011gw}.   
Heating of GSC~6214-210~B's outer disk is probably dominated by GSC~6214-210~A 
or diffuse interstellar radiation rather than irradiation by GSC~6214-210~B itself.
We therefore assume a dust temperature between 10--20~K for GSC~6214-210~B, which is
within the range of disk midplane temperatures for isolated brown dwarfs (\citealt{Ricci:2014im}).

The 3-$\sigma$ dust mass upper limit is $<$0.05~$\Mearth$ (3.9 lunar masses) for the primary and $<$0.05--0.15~$\Mearth$
(3.9--12.6 lunar masses) for the companion.  For a gas-to-dust ratio of 100, this implies putative disk masses of $<$5~$\Mearth$ 
for GSC~6214-210~A and $<$15~$\Mearth$ for B.
These results are most sensitive to the characteristic disk temperature, dipping to 0.03~$\Mearth$ (2.8 lunar masses) of dust for 
a slightly warmer temperature of 25~K, and increasing to 0.9~$\Mearth$ (76 lunar masses) for a temperature of 5~K.
The lower limit on dust temperature is set by the diffuse interstellar radiation field and cosmic rays, which even
for dense starless cloud cores that are shielded from external radiation is only about
10~K.  So although our dust mass upper limit is highly sensitive to the dust temperature, 
we consider the dust mass of $\sim$0.15~$\Mearth$ set by a $T_C$=10~K to be a fairly
strict upper limit on sub-mm-sized particles and smaller.
The total disk mass also depends on the grain size distribution, which is especially important for an older, 
presumably evolved system like this.
If planetesimals have formed then they can also lock up a significant amount of solid mass, 
though there is an upper limit to this storage since fragmenting collisions would grind them back down 
to $\sim$mm sizes that would be detectable with these ALMA observations.
Moreover, since the gas-to-dust ratio in protoplanetary disks is generally highly uncertain and evolves over time,
the total disk masses in this work are approximate at best.

\begin{figure}
  \vskip -.35 in
  \hskip -.3 in
  \resizebox{3.9in}{!}{\includegraphics{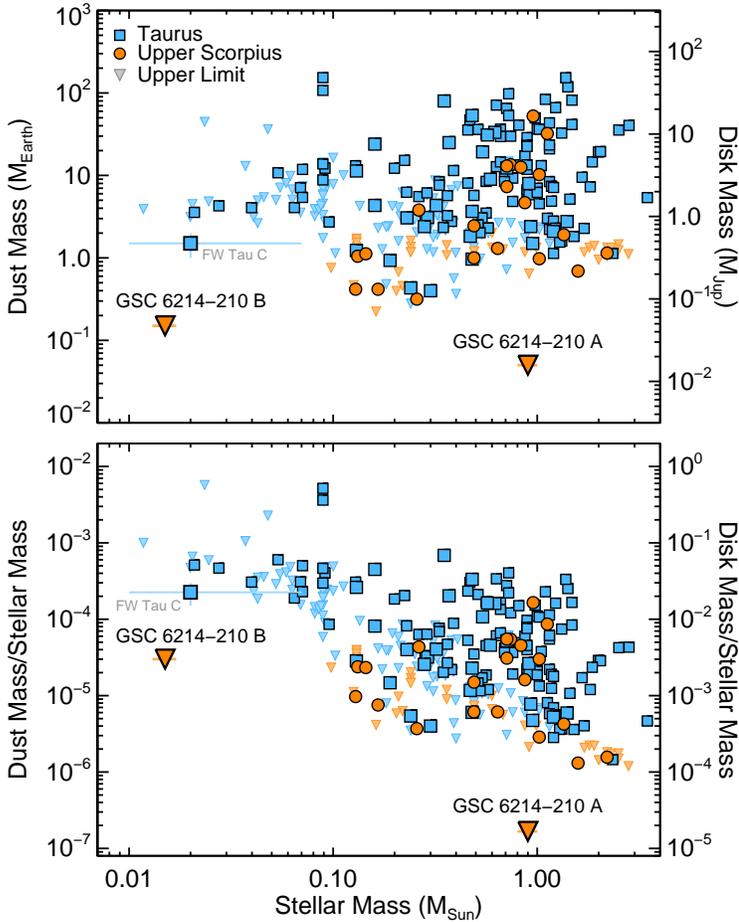}}
  \vskip -.2 in
  \caption{\emph{Top:} Sub-mm/mm dust mass measurements as a function of stellar mass for the $\approx$2~Myr Taurus (blue)
  and $\approx$5--10~Myr USco (orange) star-forming regions.  Triangles denote 3-$\sigma$ upper
  limits.  Our ALMA constraints for GSC~6214-210 A and B are over an order of magnitude lower than 
  for most previous surveys in these regions.  Taurus data are from \citet{Andrews:2013ku} and 
  \citet{Akeson:2014gp}; USco data are from
  \citet{Mathews:2012bw} and \citet{Carpenter:2014db}.  Dust masses from Andrews et al. are derived
  from their 890~$\mu$m flux density measurements and dust temperature-stellar luminosity scaling with
  a lower boundary of 10~K adopted for the lowest-luminosity brown dwarfs in their sample.  Ordinates 
  on the right-hand axis denote total disk masses assuming an interstellar-like 100:1 gas-to-dust ratio.
  For comparison, the dust mass measurement of FW~Tau~C from \citet{Kraus:2015fx} is labeled. 
  \emph{Bottom:} Similar to the upper panel, but showing dust mass to stellar mass ratio (and total disk mass to
  stellar mass ratio) as a function of stellar mass.  In terms of mass ratio, our non-detection of GSC~6214-210~B's disk
  is not particularly unusual.   \label{fig:masscomp} } 
\end{figure}

\section{Discussion}{\label{sec:intro}}

Our deep 880~$\mu$m upper limit implies that GSC~6214-210~B's disk has a low mass ($\lesssim$0.05~\Mjup) and
lacks a significant amount of cold dust in its outer regions.
This is broadly consistent with expectations from circum-planetary disk simulations, which find that subdisks are truncated at 
$\sim$1/3 R$_\mathrm{Hill}$ as a result of the gravitational influence of the host star (\citealt{Shabram:2013dm}).
For GSC~6214-210~B, this corresponds to a radius of $\approx$19~AU.  Interestingly, the
free-floating brown dwarf OTS~44 in Chamaeleon I has a similar mass ($\approx$12~\Mjup) and accretion rate 
($\sim$8$\times$10$^{-12}$ \Msun~yr$^{-1}$) to GSC~6214-210~B, but possesses a significantly
more massive disk of $\sim$30~\Mearth \ and is consistent with having a spatial extent out to $\sim$100~AU
(\citealt{Joergens:2013hw}).  These differences could be caused by disk evolution owing to the different ages
of these objects ($\sim$2~Myr versus 5--10~Myr), 
environmental influences from the nearby host star GSC~6214-210~A, natural variations 
in the physical properties of circum-substellar disks, and/or differing formation mechanisms.  
It is also possible, though more speculative, that GSC6214-210~B originally formed closer in and was dynamically scattered to
a wide separation from an as-yet-undiscovered more massive companion.  In this scenario, GSC6214-210~B's disk could have been
truncated, explaining its hot, inner component but low overall mass. 
We note that it is possible the disk may be very small ($\lesssim$0.4~AU) and optically thick given the brightness temperature limit 
of $<$0.02~K, but that would require long-term stability of the disk in the face of short viscous timescales.
Ultimately, the growing population of planetary-mass
companions being found at hundreds of AU (\citealt{Kraus:2014tl}) combined with ALMA's unprecedented sensitivity and
spatial resolution will make it possible to explore these effects for larger samples of free-floating and bound
planetary-mass objects spanning ages of 1--10~Myr.

In Figure~\ref{fig:masscomp} we compare our ALMA constraints to previous sub-mm/mm surveys in the 
Taurus and USco star-forming regions.
Our upper limit for GSC~6214-210~A indicates there is a wide range of outcomes for protoplanetary disks around Sun-like stars 
in USco.  Larger programs by \citet{Mathews:2012bw} and 
\citet{Carpenter:2014db} found disk masses in excess of 10~\Mjup, implying a spread of over 3 orders
of magnitude in disk mass when taking into account our non-detection.  
Indeed, if the primary star GSC~6214-210~A harbors a disk, it must be at most a putative 0.0015\% of the host star mass---
much lower than the median disk-to-star mass ratio of $\sim$0.3\% found at younger ages  by \citet{Andrews:2013ku}.
Altogether it is clear that giant planet formation has ceased in this system.

Similarly, there is not enough dust in GSC~6214-210~B's disk for gas giant ``moons'' to form, 
but we cannot rule out future rocky moons with masses of $\approx$0.15~$\Mearth$ ($\approx$12 lunar masses).  
\citet{Canup:2006bz} find that the shared mass ratio of Jupiter, Saturn, and Uranus to the total mass of
their regular moons ($\sim$10$^{4}$) can be explained by a balance between inflowing material
into their circum-planetary disks, which provides the raw material for creating new moons, and satellite
loss through orbit decay.  In this scenario the current regular moons around these gas giants are simply the
last generation of those formed from their subdisks.  
The current dust content of GSC~6214-210~B is a factor of at least 3 lower than this planet-moon mass ratio,
so if this phenomenon is universal then any moons already formed or in the process of forming in this system 
will probably be the last ones.  

Assuming a steady-state accretion rate, the lifetime of GSC~6214-210~B's disk can be estimated from its accretion rate 
(10$^{-10.8}$ \Msun~yr$^{-1}$; \citealt{Zhou:2014ct})
and disk mass constraint.  At most, accretion will continue for another $\sim$3 Myr, though the fractional 
gain in planet mass is only $<$0.4\%.  The GSC~6214-210~AB system therefore appears to be in its last stages of assembly,
with giant planet formation having ceased around the primary star and moon formation probably in its final
stages around the companion. \\

\acknowledgments
We are grateful to the referee for helpful comments, Jonathan Swift for productive discussions about pursuing this idea, 
and Vanessa Bailey for providing zero point flux densities for MMT and LBT filters.
This paper makes use of the following ALMA data:
ADS/JAO.ALMA\#2013.1.00487.S. ALMA is a partnership of ESO (representing
its member states), NSF (USA) and NINS (Japan), together with NRC
(Canada) and NSC and ASIAA (Taiwan), in cooperation with the Republic of
Chile. The Joint ALMA Observatory is operated by ESO, AUI/NRAO and NAOJ.
The National Radio Astronomy Observatory is a facility of the National
Science Foundation operated under cooperative agreement by Associated
Universities, Inc.
We utilized data products from the Two Micron All Sky Survey, which is a joint project of the University of 
Massachusetts and the Infrared Processing and Analysis Center/California Institute of Technology, funded by the 
National Aeronautics and Space Administration and the National Science Foundation.
 NASA's Astrophysics Data System Bibliographic Services together with the VizieR catalogue access tool and SIMBAD database 
operated at CDS, Strasbourg, France, were invaluable resources for this work.

\facility{{\it Facilities}: 
\facility{ALMA}}

%\bibliographystyle{apj}
%\bibliography{ALMA_GSC6214-210B_astroph.bbl}

\end{document}